\documentclass[preprint,amsmath,amssymb,aps,showkeys,showpacs]{revtex4}
\usepackage[english]{babel}
\usepackage{graphicx}
\usepackage{graphics}
\usepackage{amsmath}
\usepackage{dcolumn}
\usepackage{amssymb}
\usepackage{bm}

\begin{document}
\title{A relativistic quantum oscillator subject to a Coulomb-type potential induced by effects of the violation of the Lorentz symmetry}
\author{R. L. L. Vit\'oria}
\affiliation{Departamento de F\'isica, Universidade Federal da Para\'iba, Caixa Postal 5008, 58051-900, Jo\~ao Pessoa-PB, Brazil.} 

\author{H. Belich} 
\email{belichjr@gmail.com}
\affiliation{Departamento de F\'isica e Qu\'imica, Universidade Federal do Esp\'irito Santo, Av. Fernando Ferrari, 514, Goiabeiras, 29060-900, Vit\'oria, ES, Brazil.}

\author{K. Bakke}
\email{kbakke@fisica.ufpb.br} 
\affiliation{Departamento de F\'isica, Universidade Federal da Para\'iba, Caixa Postal 5008, 58051-900, Jo\~ao Pessoa-PB, Brazil.}

\begin{abstract}
We consider a background of the violation of the Lorentz symmetry determined by the tensor $\left( K_{F}\right)_{\mu\nu\alpha\beta}$ which governs the Lorentz symmetry violation out of the Standard Model Extension, where this background gives rise to a Coulomb-type potential, and then, we analyse its effects on a relativistic quantum oscillator. Furthermore, we analyse the behaviour of the relativistic quantum oscillator under the influence of a linear scalar potential and this background of the Lorentz symmetry violation. We show in both cases that analytical solutions to the Klein-Gordon equation can be achieved.

\end{abstract}

\keywords{Lorentz symmetry violation, relativistic oscillator, linear scalar potential, relativistic bound states}
\pacs{11.30.Cp, 11.30.Qc, 03.65.Pm}

\maketitle

\section{Introduction}

The size of the proton radius is an estimation given by the Standard Model (SM) and the quantum chromodynamics (QCD) of the quark interaction mediated by virtual gluons. Measurements with electrons have shown that the value of the size of the proton radius is in agreement with that yielded by the existing theories. On the other hand, recent researchers have considered a muon in orbit around a proton and shown that the radius of the proton is different, i.e., something that shouldn't have happened \cite{sci}. Such results occurred just when SM goes through its final test: the detection of the Higgs boson \cite{higgs} at the LHC in 2013. Hence, there is a fundamental question to be answered, which is the justification for the Higgs mass. Another problem related to the mass of the Higgs boson is that the energy grows beyond the SM energy scale, and thus, the radiative corrections make the mass of the Higgs boson to diverge (the problem of the Hierarchy). Besides, SM has not succeeded in explaining the origin of the electric dipole moment of the electron (EDM) \cite{revmod}. Theories that go beyond SM predict a small, but potentially measurable EDM ($d_{e}\leq 10^{-29}\,\mathrm{e}\cdot \mathrm{cm}$) \cite{science}. Experimental measurements have been improved and reached the level $d_{e}\approx 10^{-31}\,\mathrm{e}\cdot \mathrm{cm}$ \cite{Baron,Measure1,Measure2,Measure2a,Measure2b}. Despite the great success of SM to give an overview of the microscopic processes through a field theory that unifies the weak and electromagnetic interactions, SM has some limitations. In recent years, the interest in investigating the possibility of a physics beyond SM has been increased with the need of understanding Dark Matter. There is a lack of explaining Dark Matter and this has motivated new purposes of describing the interaction between the dark section and the visible sector, which could induce the detection of a weak fifth force. It is investigated in decays of an excited state of $^{8}\mathrm{Be}$ \cite{irvin}. Also, we have the unbalance between matter-antimatter that has not been clarified by the SM \cite{Axion,Axion1,Axion2,Axion3,Axion4,Axion5,Axion6}.

An interesting proposal for investigating the physics beyond SM was made by Kosteleck\'{y} and Samuel \cite{sam}, where they dealt with the spontaneous violation of symmetry through non-scalar fields (vacuum of fields that have a tensor nature) based on the string field theory. A consistent description of fluctuations around this new vacuum is obtained if the components of the background field are constant, and by the fact that this new minimum to be a background not a scalar, the Lorentz symmetry is spontaneously broken \cite{ens}. This possibility of extending SM has been considered for fields that belong to a more fundamental theory, which may induce the spontaneous violation of Lorentz symmetry based on a specific potential. It is worth mentioning that this extension of SM keeps the gauge invariance, the conservation of energy and momentum and the covariance under observer rotations and boosts, where this extension is called as the Standard Model Extension (SME) \cite{col,coll2}. In this context, it is well-known that the presence of terms that violate the Lorentz symmetry imposes at least one privileged direction in the spacetime. In recent decades, studies of the violation of the Lorentz symmetry have been made in several branches of physics \cite{h1,h2,e1,e2,e3,e4,e5,e6,e7,rasb1,w,tensor1,tensor2,geom1,geom2,geom3,bb2,bb4,book,l1,g1,g2,g3}.

In this paper, we investigate the behaviour of the Klein-Gordon oscillator \cite{kgo} in a possible scenario of anisotropy generated by a Lorentz symmetry breaking term defined by a tensor $\left( K_{F}\right)_{\mu\nu\alpha\beta}$ that corresponds to a tensor that governs the Lorentz symmetry violation out of the Standard Model Extension. The scenario of the violation of the Lorentz symmetry is determined by a non-null component of the tensor $\left( K_{F}\right)_{\mu\nu\alpha\beta}$ and a field configuration of crossed electric and magnetic field that gives rise a Coulomb-type potential. Then, we search for relativistic bound state solutions to the Klein-Gordon equation, where we show that analytical solutions can be achieved. Further, we analyse the Klein-Gordon oscillator under the influence of a linear scalar potential and the background of the Lorentz symmetry violation.

The structure of this paper is as follows: in section II, we introduce a background of the Lorentz symmetry violation defined by a tensor $\left( K_{F}\right)_{\mu\nu\alpha\beta}$ that governs the Lorentz symmetry violation out of the Standard Model Extension, and thus, establish a possible scenario of the Lorentz symmetry violation that induces a Coulomb-type potential. Then, analyse the behaviour of the Klein-Gordon oscillator under the effects of this Coulomb-type potential; in section III, we introduce a linear scalar potential as a modification of the mass term of the Klein-Gordon and investigate the effects of the linear scalar potential and the Coulomb-type potential on the Klein-Gordon oscillator; in section IV, we present our conclusions.

\section{Klein-Gordon oscillator under the effects of violation of the Lorentz symmetry}

The Klein-Gordon oscillator \cite{kgo,kgo2,kgo8,kgo7} is a relativistic model for the harmonic oscillator that interacts with scalar particles. It was proposed by Bruce and Minning \cite{kgo} in analogy with the Dirac oscillator \cite{osc1} and it has attracted interests in studies of noncommutative space \cite{kgo3,kgo4}, in noncommutative phase space \cite{kgo5}, in Kaluza-Klein theories \cite{furtado} and in $\mathcal{PT}$-symmetric Hamiltonian \cite{kgo6}. The relativistic oscillator coupling proposed by Bruce and Minning \cite{kgo} allows us the write the Klein-Gordon equation in $\left(2+1\right)$ dimensions as follows $\left(\hbar=c=1\right)$:
\begin{eqnarray}
m^{2}\phi=-\frac{\partial^{2}\phi}{\partial t^{2}}-\left[\hat{p}+im\omega\rho\,\hat{\rho}\right]\cdot\left[\hat{p}-im\omega\rho\,\hat{\rho}\right]\phi,
\label{1.1}
\end{eqnarray}
where $m$ is the rest mass of the particle, $\omega$ is the angular frequency of the Klein-Gordon oscillator, $\rho=\sqrt{x^{2}+y^{2}}$ is the radial coordinate and $\hat{\rho}$ is a unit vector in the radial direction \footnote{By following Ref. \cite{furtado}, we can write the Klein-Gordon equation (\ref{1.1}) in covariant notation by replacing $\hat{p}_{\mu}$ with $\hat{p}_{\mu}+i\,m\,\omega\,X_{\mu}$, where $X_{\mu}=\left(0,\rho,0,0\right)$. Then, the Klein-Gordon equation is written in the form: $\frac{1}{\sqrt{-g}}\left(\partial_{\mu}+m\,\omega\,X_{\mu}\right)\sqrt{-g}\,g^{\mu\nu}\left(\partial_{\nu}+m\,\omega\,X_{\nu}\right)\phi-m^{2}\phi=0$, where $g$ is the determinant of the metric tensor and $g^{\mu\nu}$ is the inverse of the metric tensor. Note that with the choice $X_{\mu}=\left(0,\rho,0,0\right)$, hence, the invariance of the Lorentz transformation of particle (or the active invariance) is broken. Indeed, this coupling preserves the invariance of the Lorentz transformation of observers as the behaviour of a genuine background field.}. Note that we have written the Klein-Gordon equation (\ref{1.1}) by using the signature of the spacetime:
\begin{eqnarray}
ds^{2}=-dt^{2}+d\rho^{2}+\rho^{2}\,d\varphi^{2}.
\label{1.2}
\end{eqnarray}

On the other hand, the gauge sector of the Standard Model Extension has two terms that modifies the transport properties of the spacetime since these terms break the Lorentz symmetry. These two terms are called as the CPT-odd sector \cite{col,coll2} and the CPT-even sector \cite{baeta,jackiw} and they start with tiny values and go for energy scales beyond the Standard Model. Recently, inspired by Refs. \cite{col,coll2,kost2,kost3,tensor1}, two of us have studied the relativistic quantum dynamics of a scalar particle under the effects of the Lorentz symmetry violation by introducing a nonminimal coupling into the Klein-Gordon equation given by $\hat{p}^{\mu}\hat{p}_{\mu}\rightarrow \hat{p}^{\mu}\hat{p}_{\mu}+\frac{g}{4}\,\left(K_{F}\right)_{\mu\nu\alpha\beta}\,F^{\mu\nu}\left(x\right)\,F^{\alpha\beta}\left(x\right)$ \cite{bb19,bb20}, where $g$ is a constant, $F^{\mu\nu}\left(x\right)$ is the electromagnetic tensor and $\left(K_{F}\right)_{\mu\nu\alpha\beta}$ corresponds to a tensor that governs the Lorentz symmetry violation out of the Standard Model Extension \cite{col,coll2,baeta,kost2,kost3}. Thereby, by introducing this nonminimal coupling into Eq. (\ref{1.1}), we have
\begin{eqnarray}
m^{2}\phi=-\frac{\partial^{2}\phi}{\partial t^{2}}-\left[\hat{p}+im\omega\rho\,\hat{\rho}\right]\cdot\left[\hat{p}-im\omega\rho\,\hat{\rho}\right]\phi+\frac{g}{4}\,\left(K_{F}\right)_{\mu\nu\alpha\beta}\,F^{\mu\nu}\left(x\right)\,F^{\alpha\beta}\left(x\right)\phi.
\label{1.3}
\end{eqnarray}
It is worth pointing out that despite being inspired by the CPT-even sector of the SME, we deal with scenarios that covers energy scales beyond the SME. For this purpose, we shall relax the renormalization property of our model. Furthermore, the study of phase transitions, which are usually described by the spontaneous symmetry breaking by a scalar field, sheds light on some phenomena in condensed matter physics and in high energy physics.	Among them, there are the problems of the baryon asymmetry, dark matter and topological defects in the early Universe. In the context of the present work, the coupling between the scalar field and the tensor field $\left(  K_{F}\right)_{\mu\nu\alpha\beta}$ describes the anisotropy of the spacetime by creating the preferred directions in the spacetime by a Lorentz symmetry violating term.

One of the properties of the tensor $\left(K_{F}\right)_{\mu\nu\alpha\beta}$ is that it can be written in terms of $3\times3$ matrices that represent the parity-even sector of it: $\left(\kappa_{DE}\right)_{jk}=-2\left(K_{F}\right)_{0j0k}$ and $\left(\kappa_{HB}\right)_{jk}=\frac{1}{2}\epsilon_{jpq}\,\epsilon_{klm}\left(K_{F}\right)^{pqlm}$; and the parity-odd sector of this tensor are represented by the $3\times3$ matrices: $\left(\kappa_{DB}\right)_{jk}=-\left(\kappa_{HE}\right)_{kj}=\epsilon_{kpq}\left(K_{F}\right)^{0jpq}$. Note that the matrices $\left(\kappa_{DE}\right)_{jk}$ and $\left(\kappa_{HB}\right)_{jk}$ are symmetric and the matrices $\left(\kappa_{DB}\right)_{jk}$ and $\left(\kappa_{HE}\right)_{kj}$ have no symmetry. In this way, we can rewrite Eq. (\ref{1.3}) in the form:
\begin{eqnarray}
m^{2}\phi&=&-\frac{\partial^{2}\phi}{\partial t^{2}}-\left[\hat{p}+im\omega\rho\,\hat{\rho}\right]\cdot\left[\hat{p}-im\omega\rho\,\hat{\rho}\right]\phi-\frac{g}{2}\left(\kappa_{DE}\right)_{i\,j}\,E^{i}\,E^{j}\,\phi\nonumber\\
[-2mm]\label{1.4}\\[-2mm]
&+&\frac{g}{2}\,\left(\kappa_{HB}\right)_{j\,k}\,B^{i}\,B^{j}\,\phi-g\left(\kappa_{DB}\right)_{j\,k}\,E^{i}\,B^{j}\,\phi.\nonumber
\end{eqnarray}

Henceforth, let us consider a possible scenario of the Lorentz symmetry violation determined by only one non-null component of the tensor $\left(\kappa_{DB}\right)_{jk}$ as being $(\kappa_{DB})_{13}=\kappa=\mathrm{const}$ and by a field configuration given by \cite{bb19}:
\begin{eqnarray}
\vec{B}=B_{0}\,\hat{z};\,\,\,\,\,\,\,\vec{E}=\frac{\lambda}{\rho}\,\hat{\rho},
\label{1.5}
\end{eqnarray}
where $B_{0}$ is a constant, $\hat{z}$ is a unit vector in the $z$-direction and $\lambda$ is a constant associated with a linear distribution of electric charges on the $z$-axis. Then, the Klein-Gordon equation (\ref{1.4}) becomes:
\begin{eqnarray}
m^{2}\phi=-\frac{\partial^{2}\phi}{\partial\,t}+\frac{\partial^{2}\phi}{\partial\rho^{2}}+\frac{1}{\rho}\frac{\partial\phi}{\partial\rho}+\frac{1}{\rho^{2}}\frac{\partial^{2}\phi}{\partial\varphi^{2}}-m^{2}\omega^{2}\rho^{2}\phi-\frac{g\lambda\,B_{0}\kappa}{\rho}\phi+m\,\omega\,\phi=0.
\label{1.6}
\end{eqnarray}
The solution to Eq. (\ref{1.6}) can be written as $\phi\left(t,\,\rho,\,\varphi\right)=\Psi\left(t\right)\,\Phi\left(\varphi\right)\,f\left(\rho\right)$; thus, we have that $\Psi\left(t\right)=e^{-i\mathcal{E}t}$, $\Phi\left(\varphi\right)=e^{il\varphi}$, where $l=0,\pm1,\pm2,\ldots$. Thereby, let us define $r=\sqrt{m\,\omega}\,\rho$, and thus, write Eq.(\ref{1.6}) in the form:
\begin{eqnarray}
\frac{d^{2}f}{dr^{2}}+\frac{1}{r}\frac{df}{dr}-\frac{l^{2}}{r^{2}}f-\frac{\alpha}{r}f-r^{2}\,f+\mu\,f=0,
\label{1.7}
\end{eqnarray}
where the parameters $\alpha$ and $\mu$ are defined as
\begin{eqnarray}
\alpha=\frac{g\,\lambda\,B_{0}\kappa}{\sqrt{m\,\omega}};\,\,\,\,\,\mu=\frac{1}{m\,\omega}\left[\mathcal{E}^{2}-m^{2}+m\,\omega\right].
\label{1.8}
\end{eqnarray}

The asymptotic behaviour of the possible solutions to Eq. (\ref{1.7}) are determined by $r\rightarrow0$ and $r\rightarrow\infty$, then, by imposing that the function $f\left(r\right)$ vanishes at $r\rightarrow0$ and $r\rightarrow\infty$, we can write the solution to Eq. (\ref{1.7}) in the form \cite{bf}:
\begin{eqnarray}
f\left(r\right)=r^{\left|l\right|}\,e^{-\frac{r^{2}}{2}}\,F\left(r\right),
\label{1.9}
\end{eqnarray}
where the function $F\left(r\right)$ is the solution to the following second-order differential equation:
\begin{eqnarray}
\frac{d^{2}F}{dr^{2}}+\left[\frac{2\left|l\right|+1}{r}-2r\right]\frac{dF}{dr}+\left[\mu-2\left|l\right|-2-\frac{\alpha}{r}\right]F=0,
\label{1.10}
\end{eqnarray}
which is called as the biconfluent Heun equation \cite{heun}, and thus, the function $F\left(r\right)$ is the biconfluent Heun function: $F\left(r\right)=H_{\mathrm{B}}\left(2\left|l\right|,\,0,\,\mu,\,2\alpha;\,r\right)$.

Let us go further by writing the function $F\left(r\right)$ as a power series expansion around the origin: $F\left(r\right)=\sum_{k=0}^{\infty}\,b_{k}\,r^{k}$, which is called as the Frobenius method \cite{arf}. By substituting this series into Eq. (\ref{1.10}), we obtain the relation
\begin{eqnarray}
b_{1}=\frac{\alpha}{2\left|l\right|+1}\,b_{0},
\label{1.11}
\end{eqnarray}
and the recurrence relation:
\begin{eqnarray}
b_{k+2}=\frac{\alpha\,b_{k+1}-\left(\mu-2\left|l\right|-2-2k\right)b_{k}}{\left(k+2\right)\left(k+2+2\left|l\right|\right)}.
\label{1.12}
\end{eqnarray}

In search of polynomial solutions to the biconfluent Heun equation (\ref{1.10}), we can see from Eq. (\ref{1.12}) that the biconfluent Heun series becomes a polynomial of degree $n$ when:
\begin{eqnarray}
\mu-2\left|l\right|-2=2n;\,\,\,\,\,\,b_{n+1}=0,
\label{1.13}
\end{eqnarray}
where $n=1,\,2,\,3\,\ldots$. Therefore, the condition $\mu-2\left|l\right|-2=2n$ yields an expression for the relativistic energy levels given by
\begin{eqnarray}
\mathcal{E}_{n,\,l}=\pm\sqrt{m^{2}+2m\omega\left[n+\left|l\right|+\frac{1}{2}\right]},
\label{1.14}
\end{eqnarray}
where $n=1,\,2,\,3\,\ldots$ and $l=0,\pm1,\pm2,\pm3,\ldots$ are the quantum numbers associated with the radial modes and the angular momentum, respectively.

However, we need to analyse the condition $b_{n+1}=0$ in order to achieve the polynomial solutions to the biconfluent Heun equation (\ref{1.10}). For this purpose, we need to know some coefficients of the power series expansion. Let us start with $b_{0}=1$, then, from Eqs. (\ref{1.11}) and (\ref{1.12}), we have that 
\begin{eqnarray}
b_{1}=\frac{\alpha}{\left(2\left|l\right|+1\right)};\,\,\,\,b_{2}=\frac{\alpha^{2}-\left(\mu-2\left|l\right|-2\right)\left(2\left|l\right|+1\right)}{2\left(2\left|l\right|+1\right)\left(2\left|l\right|+2\right)}.
\label{1.15}
\end{eqnarray}
Hence, by dealing with the lowest energy state of the system $\left(n=1\right)$, we have that $b_{n+1}=b_{2}=0$, then, we obtain the relation:
\begin{eqnarray}
\omega_{1,\,l}=\frac{\left(g\lambda\,B_{0}\,\kappa\right)^{2}}{2m\left(2\left|l\right|+1\right)},
\label{1.16}
\end{eqnarray}
which shows us that, by considering the angular frequency of the Klein-Gordon oscillator as a parameter that we can adjust, hence, some specific values of the angular frequency are permitted if we wish to achieve a polynomial solution to the function $F\left(r\right)$. In particular, Eq. (\ref{1.16}) corresponds to the possible values of the angular frequency of the Klein-Gordon oscillator associated with the lowest energy state of the system $\left(n=1\right)$ that yield a polynomial solution to the function $F\left(r\right)$, and thus, satisfy the asymptotic behaviour of the radial wave function when $r\rightarrow0$ and $r\rightarrow\infty$. Besides, we have that the possible values of the angular frequency depend on the parameters that establish the scenario of the violation of the Lorentz symmetry violation $\left(g,\,\lambda,\,\,B_{0},\,\kappa\right)$ and the quantum number of the system $\left\{n,\,l\right\}$. Therefore, the expression for the energy of the lowest energy state of the system becomes:
\begin{eqnarray}
\mathcal{E}_{1,\,l}=\pm\,m\sqrt{1+\frac{\left(g\lambda\,B_{0}\,\kappa\right)^{2}\left(2\left|l\right|+3\right)}{2m^{2}\left(2\left|l\right|+1\right)}}.
\label{1.17}
\end{eqnarray}

Finally, by labelling $\omega=\omega_{n,\,l}$, the general expression for the relativistic energy levels (\ref{1.14}) becomes:
\begin{eqnarray}
\mathcal{E}_{n,\,l}=\pm\sqrt{m^{2}+2m\omega_{n,\,l}\left[n+\left|l\right|+\frac{1}{2}\right]}.
\label{1.14a}
\end{eqnarray}

Hence, the effects of the Lorentz symmetry breaking background on the Klein-Gordon oscillator are characterized by the presence of a Coulomb-type potential that modifies the spectrum of energy and imposes a restriction on the possible values of the angular frequency of the Klein-Gordon oscillator. This restriction shows us that the parameters associated with the background of the violation of the Lorentz symmetry and the quantum numbers of the system determine the allowed values of the angular frequency which yield a polynomial solution to the function $F\left(r\right)$.

\section{Klein-Gordon oscillator under the effects of violation of the Lorentz symmetry and a linear scalar potential}

In this section, we analyse the relativistic quantum effects of a linear scalar potential and the violation of the Lorentz symmetry on the Klein-Gordon oscillator. The linear scalar potential is introduced in the Klein-Gordon equation (\ref{1.4}) through the modification of the mass term as \cite{greiner,vb2}
\begin{eqnarray}
m\left(\rho\right)=m+\chi\,\rho,
\label{2.1}
\end{eqnarray}
where $\chi$ is a constant that characterizes the linear confining potential. It is worth mentioning the great interest of the linear scalar potential in quantum field theory \cite{linear,linear1,linear4,linear4a,linear4b,linear4c}.  Therefore, by considering the background of the violation of the Lorentz symmetry of the previous section, then, the Klein-Gordon equation (\ref{1.6}) becomes
\begin{eqnarray}
\left(m+\chi\,\rho\right)^{2}\phi=-\frac{\partial^{2}\phi}{\partial\,t}+\frac{\partial^{2}\phi}{\partial\rho^{2}}+\frac{1}{\rho}\frac{\partial\phi}{\partial\rho}+\frac{1}{\rho^{2}}\frac{\partial^{2}\phi}{\partial\varphi^{2}}-m^{2}\omega^{2}\rho^{2}\phi-\frac{g\lambda\,B_{0}\kappa}{\rho}\phi+m\,\omega\,\phi=0.
\label{2.2}
\end{eqnarray}

By following the steps from Eq. (\ref{1.6}) to Eq. (\ref{1.8}), we can define the parameter
\begin{eqnarray}
\varpi^{2}=m^{2}\omega^{2}+\chi^{2},
\label{2.3}
\end{eqnarray}
and make a change of variable given by $\zeta=\sqrt{\varpi}\,\rho$; thus, we obtain
\begin{eqnarray}
\frac{d^{2}f}{d\zeta^{2}}+\frac{1}{\zeta}\frac{df}{d\zeta}-\frac{l^{2}}{\zeta^{2}}f-\frac{\theta}{\zeta}f-\zeta^{2}\,f-\delta\,\zeta\,f+\bar{\mu}\,f=0,
\label{2.4}
\end{eqnarray}
where the new set of parameters is defined as
\begin{eqnarray}
\theta=\frac{g\,\lambda\,B_{0}\kappa}{\sqrt{\varpi}};\,\,\,\delta=\frac{2m\chi}{\varpi^{3/2}};\,\,\,\bar{\mu}=\frac{1}{\varpi}\left[\mathcal{E}^{2}-m^{2}+m\,\omega\right].
\label{2.5}
\end{eqnarray}

Let us impose that the function $f\left(\zeta\right)$ vanishes at $\zeta\rightarrow0$ and $\zeta\rightarrow\infty$, then, we can write the solution to Eq. (\ref{2.5}) in the form \cite{vb2}:  
\begin{eqnarray}
f\left(\zeta\right)=\zeta^{\left|l\right|}\,e^{-\frac{\zeta^{2}}{2}}\,e^{-\frac{\delta\,\zeta}{2}}\,F\left(\zeta\right),
\label{2.6}
\end{eqnarray}
where the function $F\left(\zeta\right)$ is the solution to the following equation:
\begin{eqnarray}
\frac{d^{2}F}{d\zeta^{2}}+\left[\frac{2\left|l\right|+1}{\zeta}-\delta-2\zeta\right]\frac{dF}{d\zeta}+\left[\bar{\mu}+\frac{\delta^{2}}{4}-2\left|l\right|-2-\frac{2\theta+\delta\left(2\left|l\right|+1\right)}{2\zeta}\right]F=0,
\label{2.7}
\end{eqnarray}
which is also the biconfluent Heun equation \cite{heun} and the function $F\left(\zeta\right)$ is also the biconfluent Heun function, which is determined by: $F\left(\zeta\right)=H_{\mathrm{B}}\left(2\left|l\right|,\,\delta,\,\bar{\mu}+\frac{\delta^{2}}{4},\,2\alpha;\,\zeta\right)$.

Let us proceed with the Frobenius method \cite{arf} as we have made from Eq. (\ref{1.11}) to Eq. (\ref{1.13}). In the present case, we must write $F\left(\zeta\right)=\sum_{k=0}^{\infty}\,b_{k}\,\zeta^{k}$, and then, we obtain the relation
\begin{eqnarray}
b_{1}=\frac{\delta\left(2\left|l\right|+1\right)-2\theta}{2\left(2\left|l\right|+1\right)}\,b_{0},
\label{2.8}
\end{eqnarray}
and the recurrence relation:
\begin{eqnarray}
b_{k+2}=\frac{\left[\delta\left(2k+2\left|l\right|+3\right)-2\theta\right]\,b_{k+1}-\left(2\bar{\mu}+\frac{\delta^{2}}{2}-4\left|l\right|-4-4k\right)b_{k}}{2\left(k+2\right)\left(k+2+2\left|l\right|\right)}.
\label{2.9}
\end{eqnarray}

In search of polynomial solutions to the biconfluent Heun equation (\ref{1.10}), we can see from Eq. (\ref{1.12}) that the biconfluent Heun series becomes a polynomial of degree $n$ when:
\begin{eqnarray}
2\bar{\mu}+\frac{\delta^{2}}{2}-4\left|l\right|-4=4n;\,\,\,\,\,\,b_{n+1}=0,
\label{2.10}
\end{eqnarray}
where $n=1,\,2,\,3\,\ldots$. Therefore, the condition $2\bar{\mu}+\frac{\delta^{2}}{2}-4\left|l\right|-4=4n$ yields an expression for the relativistic energy levels given by
\begin{eqnarray}
\mathcal{E}_{n,\,l}=\pm\sqrt{m^{2}+2\sqrt{m^{2}\omega^{2}+\chi^{2}}\left[n+\left|l\right|+1\right]-m\omega-\frac{m^{2}\chi^{2}}{m^{2}\omega^{2}+\chi^{2}}},
\label{2.11}
\end{eqnarray}
where $n=1,\,2,\,3\,\ldots$ and $l=0,\pm1,\pm2,\pm3,\ldots$ are the quantum numbers associated with the radial modes and the angular momentum, respectively.

Before analysing the condition $b_{n+1}=0$ given in Eq. (\ref{2.10}), let us consider the first coefficient of the power series expansion $F\left(\zeta\right)=\sum_{k=0}^{\infty}\,b_{k}\,\zeta^{k}$ as being $b_{0}=1$, and thus we have
\begin{eqnarray}
b_{1}&=&\frac{\delta\left(2\left|l\right|+1\right)-2\theta}{2\left(2\left|l\right|+1\right)};\nonumber\\
[-2mm]\label{2.12}\\[-2mm]
b_{2}&=&\frac{\delta^{2}\left(2\left|l\right|+3\right)}{8\left(2\left|l\right|+2\right)}-\frac{\theta\delta\left(\left|l\right|+1\right)}{\left(2\left|l\right|+2\right)\left(2\left|l\right|+1\right)}+\frac{\theta^{2}}{2\left(2\left|l\right|+2\right)\left(2\left|l\right|+1\right)}-\frac{h}{4\left(2\left|l\right|+2\right)},\nonumber
\end{eqnarray}
where $h=2\bar{\mu}+\frac{\delta^{2}}{2}-4\left|l\right|-4$. Hence, by taking into account the lowest energy state of the system $\left(n=1\right)$, we have that $b_{n+1}=b_{2}=0$, then, we obtain 
\begin{eqnarray}
\left(m^{2}\omega^{2}_{1,\,l}+\chi^{2}\right)^{3/2}&-&\frac{\left(g\lambda\,B_{0}\,\kappa\right)^{2}}{2\left(2\left|l\right|+1\right)}\,\left(m^{2}\omega^{2}_{1,\,l}+\chi^{2}\right)+\frac{2m\chi\,g\lambda\,B_{0}\,\kappa\left(\left|l\right|+1\right)}{2\left|l\right|+1}\,\left(m^{2}\omega^{2}_{1,\,l}+\chi^{2}\right)^{1/2}\nonumber\\
[-2mm]\label{2.13}\\[-2mm]
&-&\frac{m^{2}\chi^{2}}{2}\left(2\left|l\right|+3\right)=0,\nonumber
\end{eqnarray}
where we have that the possible values of the angular frequency of the Klein-Gordon oscillator, which yield a polynomial solution to the function $F\left(\zeta\right)$, are determined by the real solution to Eq. (\ref{2.13}). In particular, the real solution to Eq. (\ref{2.13}) is very long, therefore, we do not write it. We can also see, from Eq. (\ref{2.13}), that the possible values of the angular frequency of the Klein-Gordon oscillator depend on the quantum numbers of the system and the parameters associated with the background of the Lorentz symmetry violation. Therefore, for each relativistic energy level, we have a different relation of the angular frequency of the Klein-Gordon oscillator to the parameters of the violation of the Lorentz symmetry and the quantum numbers of the system. Then, by labelling $\omega=\omega_{n,\,l}$, we rewrite Eq. (\ref{2.12}) as  
\begin{eqnarray}
\mathcal{E}_{n,\,l}=\pm\sqrt{m^{2}+2\sqrt{m^{2}\omega^{2}_{n,\,l}+\chi^{2}}\left[n+\left|l\right|+1\right]-m\omega_{n,\,l}-\frac{m^{2}\chi^{2}}{m^{2}\omega^{2}_{n,\,l}+\chi^{2}}}.
\label{2.14}
\end{eqnarray}

Thereby, the presence of the linear scalar potential and the Coulomb-type potential induced by the effects of the Lorentz symmetry violation background modifies the spectrum of energy of the Klein-Gordon oscillator and also imposes a restriction on the values of the angular frequency of this relativistic quantum oscillator. Therefore, for each relativistic energy level, the relation of the angular frequency of this relativistic quantum oscillator to the parameters of the violation of the Lorentz symmetry and the quantum numbers of the system is different. In particular, we have seen that the possible values of the angular frequency associated with the ground state of the system are determined by Eq. (\ref{2.13}), and thus, a polynomial solution to the function $F\left(\zeta\right)$ is achieved.

\section{Conclusions}

We have analysed the behaviour of the Klein-Gordon oscillator \cite{kgo,kgo2,kgo8,kgo7,kgo9}, under the effects of a Coulomb-type potential induced by the effects of the violation of the Lorentz symmetry. We have seen that the relativistic energy levels of this relativistic quantum oscillator are modified by the background of the Lorentz symmetry violation, and also that there is a restriction on the possible values of the angular frequency of the relativistic quantum oscillator in order that a polynomial solution to the Heun biconfluent series can be achieved. This restriction shows us that, for each relativistic energy level, there is the relation of the angular frequency of this relativistic quantum oscillator to the parameters of the violation of the Lorentz symmetry and the quantum numbers of the system. In particular, we have obtained the exact expression for the angular frequency and the energy levels associated with the lowest energy state of the system.

Besides, we have considered the presence of a linear scalar potential, which has been introduced into the Klein-Gordon equation as a modification of the mass term, then, we have shown that the quantum effects that stem from the influence of the linear scalar potential and the Coulomb-type potential induced by the effects of the Lorentz symmetry violation background modify the spectrum of energy of the Klein-Gordon oscillator and also restrict the values of the angular frequency of this relativistic quantum oscillator in order to obtain a polynomial solution to the Heun biconfluent series. We have also seen that for each energy levels there is a different relation of the angular frequency of this relativistic quantum oscillator to the parameters of the violation of the Lorentz symmetry and the quantum numbers of the system. By analysing the lowest energy state of the system, we have shown that the permitted values of the angular frequency of the relativistic quantum oscillator are determined by an algebraic equation.

\acknowledgments{The authors would like to thank the Brazilian agencies CNPq and CAPES for financial support.}

\end{document}